\documentclass[12pt]{iopart}
\usepackage{graphicx}
\usepackage{bm}
\usepackage{mathrsfs}

\begin{document}
\date{\today}

\title{Open End Correction for Flanged Circular Tube from the Diffusion Process}

\author{Naohisa Ogawa}
\ead{ogawanao@hit.ac.jp}
\address{Hokkaido Institute of Technology, 7-15 Maeda, Teine Sapporo 006-8585 Japan.}
\author{Fumitoshi Kaneko}
\ead{toshi@chem.sci.osaka-u.ac.jp}
\address{Department of Macromolecular Science, Graduate School of Science, 
Osaka university, Toyonaka Osaka 560-0043 Japan.}

\begin{abstract}
In the physics lessons on waves and resonance phenomena in high school and college, 
we usually consider sound waves in a tube with open or closed ends \cite{ResnikHalliday}.
However, it is well known that we need a tube with open end correction  $\Delta L$. The correction for a flanged circular tube was first given by Rayleigh \cite{Rayleigh} and experimentally checked by several authors \cite{exp}.  In this paper, we show the different methods of  obtaining the end correction for a circular tube by a diffusion process.
\end{abstract}
\pacs{*43.40.Le, *43.20.Rz, 01.40.-d, 51.20.+d}
\vspace{2pc}
\submitto{\EJP}
\maketitle

\section{Introduction}

A standing sound wave in a tube with an open end has an antinode bit outside the end point.
Such a small distance is called the open end correction (hereafter abbreviated as OEC).
 The end correction $\Delta L$ was calculated for an infinite flanged circular tube by Rayleigh \cite{Rayleigh}, and after that, H. Levine and J. Schwinger calculated the unflanged end correction as a function of wavelength \cite{Schwinger}. 
OEC  for an infinite flanged circular tube as a function of wavelength was determined by Y. Nomura et al. by the radiation impedance method  \cite{Nomura}.
The recent application of OEC was given by M. S. Howe \cite{Howe} (and related references therein) .
 
 The idea of obtaining  OEC as discussed by Rayleigh comes from the fluid mechanical consideration \cite{Rayleigh}, \cite{Howe}.  Let us suppose a half infinite space with a connected circular tube filled with nonrotational and noncompressive fluid, and assume a piston moving at a speed of  $V$ in the tube as seen in fig. 1.  Then, the total kinetic energy of the fluid is given by the definition of OEC.
 
 \begin{figure}
\centerline{\includegraphics[width=3cm]{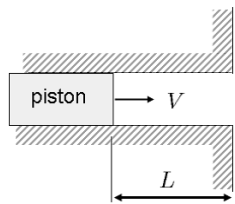}}
\caption{Moving piston in a tube connected to half infinite space filled with noncompressive fluid.}
\end{figure}
 
 \begin{equation}
 \frac{1}{2} \rho \int (\vec{\nabla} \phi)^2 d^3x  = \frac{1}{2} \rho V^2 S (L + \Delta L),
 \end{equation}
 where $\rho$ is the fluid mass density, $\phi$ is the velocity potential, $S$ is the cross-sectional area of the tube, and $\Delta L$ is the OEC. 
The integration region is taken toward the piston inner circular tube and half infinite 3D space.
The region with a nonzero velocity is not only in the tube but is also spread around a branching bay with a speed lower than $V$. Such a region contributes to OEC.

In this study, we calculate OEC quite differently.
As shown in fig. 2, we place a circular well at $x=y=0$ with a radius $a$ and a depth $L$, 
and consider the particle diffusion of density $n(x,y,z)$.
We put  a particle bath with a density $n_s$ at the bottom of the well; we thus have $n=0$ at infinity. By using the above boundary conditions, we approximately solve the static diffusion equation under the condition that one-dimensional diffusion is occurring in the well.
By using a solution for a diffusion field, we define OEC and show that its value is $8a/3\pi$, consistent with that  given by Rayleigh and Nomura et al., when the tube radius is small enough compared with the wavelength $a/\lambda <<1$ \cite{Nomura}.

\section{Estimation of diffusion in a well}
In the well, we approximate the diffusion process by using a one-dimensional diffusion equation  (see fig. 2).
This is very simple, but is essential for our further discussion.
At the top of the well, the space is widely opening and it is therefore possible to suppose that 
 $n_{top}\equiv n(z=0, r<a) =0$.
Then, the constant diffusion flow comes up from the bottom.
\begin{equation}
n(z) = n_s - \frac{J}{D} (z +L), ~~J =  \frac{n_s}{L/D}. \label{eq:innerwell_1}
\end{equation}
Our purpose is to correct this flow formula.
\\
We take $n_{top} \neq 0$ and then obtain the diffusion field and constant diffusion flow.
\begin{equation}
n(z) = n_s - \frac{J'}{D} (z +L), ~~J' =  \frac{n_s-n_{top}}{L/D}. \label{eq:innerwell}
\end{equation}
$\pi a^2 J'$ is the number of particles supplied per unit time from the well 
to the open region while leaving $n_{top}$ undetermined.
Then, we solve the diffusion equation in open space with the boundary conditions

\begin{eqnarray}
n(r, z=\infty) = n(r=\infty,z>0)=0, \label{eq:BC1} 
\end{eqnarray}

\begin{eqnarray}
\frac{\partial n}{\partial z} =\left\{\begin{array}{lll}
0 & \cdots & r \ge a, z=0\\
-J'/D & \cdots & r<a, z=0.\\
\end{array}\right. \label{eq:BC2}
\end{eqnarray} 

The additional condition 
\begin{equation}
n(z=0, r <a) = n_{top} \label{eq:BC4}
\end{equation}
 is given below using the solution of the diffusion equation in an open region.
Then, two diffusion fields in the inner and outer wells are connected, and the diffusion equation is completely solved.

\section{Diffusion in open region}
We consider the diffusion equation $\Delta n =0$ in the open region $z\ge 0$.
The boundary conditions are (\ref{eq:BC1}) and  (\ref{eq:BC2}).

\begin{figure}
\centerline{\includegraphics[width=6cm]{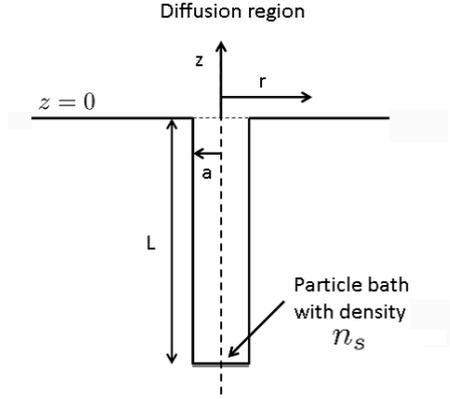}}
\caption{Circular well with depth $L$. The particle server is at the bottom. Particles diffuse from the bottom of the well into the outer region.}
\end{figure}

By supposing an axial symmetry, we obtain the following static diffusion equation for the diffusion field $n$:

\begin{equation}
\frac{\partial^2 n}{\partial r^2}  + \frac{1}{r}\frac{\partial n}{\partial r} +\frac{\partial^2 n}{\partial z^2} =0.
\end{equation}
Then, we have a solution in the form

\begin{equation}
n(r,z) = \int_0^\infty ~d\lambda ~f_{\lambda} ~J_0(\lambda r) ~e^{-\lambda z}.
\end{equation}
On the plane $z=0$, the diffusion flow in the $+z$ direction is given by

\begin{equation}
J(z=0) = -D \partial_z n \mid_{z=0} = D \int_0^\infty \lambda ~f_{\lambda} ~J_0(\lambda r) ~d\lambda. \label{eq:flow}
\end{equation}

$f_\lambda$ should be selected to obtain the boundary condition  (\ref{eq:BC2}).

The following formula solves the problem.
\begin{eqnarray}
\int_0^\infty J_1(a\lambda) J_0(r \lambda) ~ d\lambda =\left\{ \begin{array}{lll}
1/a & \cdots & (a>r>0)\\
 0 & \cdots & (r>a>0).\\
\end{array} \right. \label{eq:formula}
\end{eqnarray}

From (\ref{eq:BC2}), (\ref{eq:flow}), and (\ref{eq:formula}), we determine the function $f_\lambda$ given as
\begin{equation}
f_{\lambda} = \frac{aJ'}{D \lambda} J_1(a \lambda).
\end{equation}

Then, we obtain the following solution for $n$:

\begin{equation}
n(r,z) = \frac{aJ'}{D} \int_0^\infty \frac{d\lambda}{\lambda} J_1(a \lambda) J_0(\lambda r) e^{-\lambda z}. \label{eq:density}
\end{equation}

Next, we must determine $n_{top} = n(r<a, z=0)$.
From (\ref{eq:density}), we obtain
\begin{equation}
n(r,0) = \frac{aJ'}{D}\int_0^\infty \frac{dx}{x} J_1(x) J_0(\frac{r}{a} x) \equiv \frac{aJ'}{D} N(r/a).
\end{equation}

\begin{figure}
\centerline{\includegraphics[width=7cm]{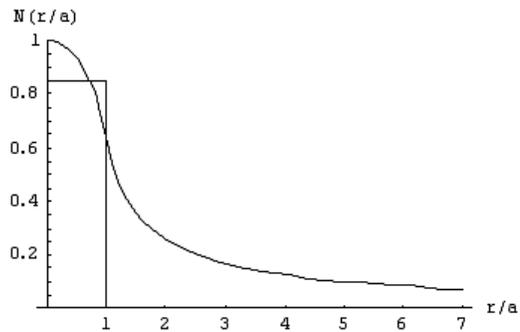}}
\caption{Density distribution on surface $z=0$. 
At $r>a$, we have a nonzero density. 
This shows that particles diffused from the well and distributed around it.
The square line shows the mean  $N(r/a)~ \mbox{for}~ r<a$.}
\end{figure}

The function $N(r/a)$ is expressed by the following hypergeometric function:
\begin{eqnarray}
N(r/a) =\left\{ \begin{array}{lll}
F(1/2, -1/2, 1; (r/a)^2)& \cdots & r<a\\
 2/\pi & \cdots & r=a\\
 F(1/2, 1/2, 2; (a/r)^2)/ (2r/a) & \cdots & r>a.\\
\end{array} \right.
\end{eqnarray}
The form of this function is given in fig. 3.
The smoothed curve is the function $N(r/a)$. 
To obtain a consistent solution with a diffusion field in the well, the matching condition (\ref{eq:BC4}) is required at $(r<a, z=0)$.
However, as shown in figure 3, the distribution at $r/a<1$ is not constant. This shows that our solution (\ref{eq:density}) is not consistent with the inner-well solution, since our approximated solution (\ref{eq:innerwell}) counts the diffusion out  in the $r$-direction.  To further achieve what in this approximation, we approximate the form of $N(r/a)$ by its mean in the region $r/a<1$.

\begin{equation}
<N> = \frac{2}{a^2} \int_0^a dr r F(1/2, -1/2, 1; (r/a)^2) \equiv K = \frac{8}{3\pi} \sim 0.8488 .
\end{equation}

This mean is shown by a solid square line in fig. 3.
Then, an additional boundary condition (\ref{eq:BC4}) leads to
\begin{equation}
n_{top} = \frac{aK}{D}J'. \label{eq:boundary1}
\end{equation}

From (\ref{eq:innerwell}) and (\ref{eq:boundary1}), we obtain

\begin{equation}
J'=\frac{n_s}{(L+aK) /D}. \label{eq:oec1}
\end{equation}

This result should be compared with the second equation of  (\ref{eq:innerwell_1}); it shows that 
the effective depth of the well is slightly larger than $L$ and that OEC is defined by

\begin{equation}
\Delta L = aK = \frac{8a}{3\pi}. \label{eq:oec2}
\end{equation}

This is the same OEC for sound in a long-wavelength limit \cite{Nomura}.
The point $z= \Delta L$ upwards of the well is the place where the density $n$ vanishes effectively.

Note that, from eq. (\ref{eq:boundary1}) OEC is defined by
\begin{equation}
\Delta L = \frac{<n>}{J'/D}, \label{eq:oec3}
\end{equation}
where
$$ <n> = \frac{1}{\pi a^2} \int_{r<a} n (r, z=0) ~dS.$$

\section{Discussion}
Why do we have OEC for sound in a long-wavelength limit by our method?
We discuss this point and show that the definition of OEC by the radiation impedance method 
is quite similar to that given in this paper.

First we review the radiation impedance method of obtaining OEC.
Let us put a disk as a sound source oscillating in the $z$ direction and having a radius $a$ on the  surface $z=0$. 
The sound emitted from the source satisfies the wave equation,

\begin{equation}
(\Delta - \frac{1}{c^2} \frac{\partial^2}{\partial t^2} ) p =0, \label{eq:wave}
\end{equation}
where $p$ is the pressure field and $c$ is the speed of sound.

The linearized Euler equation is necessary to consider the speed of sound source:
\begin{equation}
\rho \frac{\partial \vec{u}}{\partial t} = - \vec{\nabla} p, \label{eq:euler}
\end{equation}
where $\rho$ is the mass density.

From the solution of eq. (\ref{eq:wave}) and (\ref{eq:euler}),
we obtain the fluid velocity $\vec{u}$.

The boundary conditions for $p$ at the surface $z=0$ are as follows:

\begin{eqnarray}
\frac{\partial p }{\partial z} =\left\{\begin{array}{lll}
0~(u_z =0) & \cdots & r \ge a,~z=0,\\
-i \omega ~\rho ~u_{z ~(\mbox{source})} & \cdots & r < a,~z=0,\\
\end{array}\right. \label{eq:BC5}
\end{eqnarray} 
where the time dependence is supposed to be $\exp i \omega t$.

From the emitted sound $p(\vec{x},t)$, we obtain the force of reaction $F$ to the sound source as

\begin{equation}
F = \int_{r<a}  p_{(z=0)} dS.
\end{equation}
Then, the radiation impedance is defined by
\begin{equation}
Z = \frac{F}{u_{z ~ (\mbox{source})}}= -i \rho \omega \frac{\int_{r<a}  p_{(z=0)} dS}{\partial_z p_{(r<a)}}.
\end{equation}
From the parallelism to the mechanical impedance, the additional inertial mass due to the radiation is defined using the maginary part of radiation impedance.
\begin{equation}
\delta m \equiv \frac{1}{\omega} \Im(Z).
\end{equation}

This additional mass is supplied by OEC as

\begin{equation}
\delta m = \rho~ \Delta L ~ (\pi a^2).
\end{equation}

So, we obtain OEC by using the formula

\begin{equation}
\Delta L = \frac{\Im(Z)}{\rho \pi a^2 \omega} =  - \Re( \frac{<p>}{\partial_z p_{(r<a)}}) , \label{eq:oec4}
\end{equation}
where
$$ <p> = \frac{1}{\pi a^2}\int_{r<a}  p_{(z=0)} dS.$$

Now, let us check the similarity of this OEC in a diffusion process.
The long-wavelength limit is given as

$$ \lambda \to \infty, ~~\omega \to 0.$$

Then eq. (\ref{eq:wave}) leads to the Laplace equation.
The linearized Euler equation (\ref{eq:euler}) should be compared with the equation of a diffusion flow.

$$ \vec{J}/D = - \vec{\nabla} n. $$

Therefore, we have the correspondences; $n ~\sim ~p$ and $\vec{J}/D ~ \sim ~i \omega \rho \vec{u}$.
The boundary condition (\ref{eq:BC5}) corresponds to
\begin{eqnarray}
\frac{\partial n}{\partial z} =\left\{\begin{array}{lll}
0 & \cdots & r \ge a, ~z=0,\\
-J/D & \cdots & r<a,~z=0,\\
\end{array}\right. \label{eq:BC6}
\end{eqnarray} 
which is the same as (\ref{eq:BC2}).

The definition of OEC  as eq. (\ref{eq:oec4}) corresponds to
\begin{equation}
\Delta L =  - \frac{<n>}{\partial_z n_{(r<a)}} = \frac{<n>}{J/D}.
\end{equation}
This is the same as the definition of OEC for the diffusion equation (\ref{eq:oec3}).
Therefore, the equations, boundary conditions, and definitions of OEC are the same in the long-wavelength limit of these two methods. 
This is the reason why we can calculate OEC of sound waves simply using a diffusion equation.
The difficulty in obtaining OEC is quite simplified in the long-wavelength limit by considering the diffusion equation.

\section{Acknowledgments}
The authors would like to thanks Prof.Yokoyama in Gakushuin university for the helpful suggestion.

\end{document}